\documentclass[epj,nopacs]{svjour}
\usepackage{ulem}
\usepackage{amssymb}
\usepackage{amsmath}
\usepackage{epsfig}
\usepackage{color}
\usepackage{graphics, graphicx}
\usepackage{bbold}
\usepackage{psfrag}
\usepackage{mathcomp}
\usepackage{subfigure}
\usepackage{verbatim}
\usepackage{xcolor}
\usepackage[colorlinks,citecolor=blue,urlcolor=black]{hyperref}
\usepackage{cite}

\usepackage{mathrsfs}

\begin{document}

\date{\today}
\title{Interaction-modulated tunneling dynamics in a mixture of Bose-Einstein condensates}
\author{Mudassar Maraj\inst{1,}\thanks{\emph{e-mail:} mudassar.maraj@gmail.com} \footnotemark[2],Jing-Bo Wang\inst{2,3,}\thanks{These authors contributed equally.},Jian-Song Pan\inst{2,3},Wei Yi\inst{2,3}}
\institute{Shanghai Branch, National Laboratory for Physical Sciences at Microscale and Department of Modern
Physics, University of Science and Technology of China, Hefei, Anhui 230026, China \and Key Laboratory of Quantum Information, University of Science and Technology of China, Chinese Academy of Sciences, Hefei, Anhui 230026, China \and Synergetic Innovation Center of Quantum Information and Quantum Physics, University of Science and Technology of China, Hefei, Anhui 230026, China}

\date{Received: date / Revised version: date}

\abstract{
We study the interaction-modulated tunneling dynamics of a Bose-Einstein condensate (BEC) in a deep double-well potential, where the tunneling between the two wells is modulated by another BEC trapped in a harmonic potential symmetrically positioned at the center of the double-well potential. The inter-species interactions couple the dynamics of the two BECs, which give rise to interesting features in the tunneling oscillations. Adopting a two-mode approximation for the BEC in the double-well potential and coupling it with the Gross-Pitaevskii equation of the harmonically trapped BEC, we numerically investigate the coupled dynamics of the BEC mixture, and map out the phase diagram of the tunneling dynamics. We show that the dynamical back action of the BEC in the harmonic trap leads to strong non-linearity in the oscillations of the BEC in the double-well potential, which enriches the system dynamics, and enhances macroscopic self trapping. The transition between the Josephson oscillation and the self-trapping dynamics can be identified by monitoring the oscillation frequency of the double-well BEC. Our results suggest the possibility of tuning the tunneling dynamics of BECs in double-well potentials.}
\PACS{67.85.Lm, 03.75.Ss, 05.30.Fk}
\authorrunning{Mudassar Maraj,Jing-Bo Wang,Jian-Song Pan \and Wei Yi}
\titlerunning{Tunneling dynamics in a mixture of BECs}
\maketitle

\section{Introduction}

Tunneling dynamics of BECs in double-well potentials have been extensively studied ever since the experimental realization of Bose-Einstein condensate (BEC) in cold atomic gases~\cite{experiment4,experiment5,experiment10,experiment11,experiment12,spinBEC,experiment9,twoBECinDW}. This is motivated by the analogy between BECs with two spatial modes separated by a potential barrier~\cite{quantumphase,oneBECinDW2,oneBECinDW1,BoseEinsteincondensation}, and the Josephson junctions in superconductors~\cite{Superconduct,Superconduct1,Superconduct2}. Whereas the superconductor Josephson junction can be described in terms of a rigid pendulum, the tunneling dynamics of a BEC in double-well potentials can be viewed as a non-rigid pendulum~\cite{Quantumcoherent,Josephsoneffects,Bosecondensatetunneling}, which originates from superfluid-density oscillations and the density-dependent interactions. Previous theoretical and experimental studies have revealed rich phase-space dynamics in the tunneling oscillations of BECs and Fermi superfluid in BEC crossover in double-well potentials~\cite{GPEbeyond,twomode1,twomode2,twomode3,symmetrybreaking1,symmetrybreaking2,Twomodeeffective,Dynamicsinasymmetric,Phaseseparation,Symmetrybreaking,Dynamicsofcomponent,Ultracoldbosons}. These include zero- and $\pi$-phase oscillations, as well as the macroscopic quantum self-trapping (MQST) of atoms~\cite{Coherentoscillation}. Based on these understandings, recent experimental progresses in cold atomic gases have further enabled the study of tunneling dynamics in related systems~\cite{experiment1,experiment2,experiment3,experiment7,experiment8}. For instance, the experimental realization of spinor BECs has lead to the investigation of internal-state oscillations, which demonstrate similar phase-space dynamics in terms of non-rigid pendulum motion~\cite{Phaseseparation3,Numberfluctuations}. All these findings have extended the study of tunneling dynamics in superconductor Josephson junctions, and have enriched our understandings of macroscopic quantum matter.

In this work, we consider the interaction-modulated tunneling dynamics in a mixture of Bose-Einstein condensates, in which the oscillation of a BEC in a deep double-well potential is modulated by another BEC trapped in a harmonic potential symmetrically positioned at the center of the double-well potential. With the recent experimental realization of superfluid mixtures in cold atomic gases, it is promising to implement the setup experimentally using BEC mixtures loaded in optical lattice potentials. In such a system, the modulation of the tunneling dynamics derives from the inter-species interactions, which couple the dynamics of the two BECs. Affected by the tunneling oscillation of the BEC in the double-well potential, the harmonically trapped BEC also undergoes oscillations, the back action of which further modulates the tunneling dynamics of the BEC in the double well. To capture the dynamic back actions in the BEC mixture, we describe the harmonically trapped BEC with time-dependent Gross-Pitaevskii (GP) equation, while treating the BEC in the double-well potential using the conventional two-mode approximation. Compared to the more accurate approach of treating both BECs with GP equations, we show that our treatment should be accurate when the inter-species interaction energy is not too large.

A key property of the tunneling dynamics in the current system is the highly non-linear behavior. Whereas the dynamics would be reduced to the case of Josephson-like tunneling dynamics in the limits of vanishingly small or overwhelmingly large atom numbers in the harmonically trapped BEC, the tunneling dynamics is strongly modified when the atom numbers of the two BECs are comparable. For the simple tunneling dynamics in the absence of the harmonically trapped BEC, the phase-space dynamics follow one of the following modes: zero- or $\pi$-phase oscillation, where the average atom-number difference between the two wells is zero and the average relative phase is zero or $\pi$, respectively; the MQST, where the average atom-number difference is finite. In the presence of the harmonically trapped BEC, we find that, while the zero- and the $\pi$-phase oscillations can still take place over a wide parameter regime, the MQST is greatly enhanced. Our findings suggest the possibility of tuning the tunneling dynamics of BECs in double-well potentials.

The paper is organized as follows. In Sec. 2, we introduce the model Hamiltonian as well as our mean-field approach. We present the main results on the coupled dynamics in Sec. 3. In Sec. 4, we map out the phase diagram of the system dynamics. We then demonstrate in Sec. 5 that the transition between the Josephson oscillation and the self-trapping dynamics can be identified by monitoring the oscillation frequency of the double-well BEC. In Sec. 6, we compare results with the coupled GP equations approach, and discuss the validity of the two-mode approximation for the description of the BEC in the double-well potential. Finally, we summarize in Sec. 7.

\section{Model}
\begin{figure}[ht]
  \centering
  \includegraphics[width=5.5cm,height=4cm]{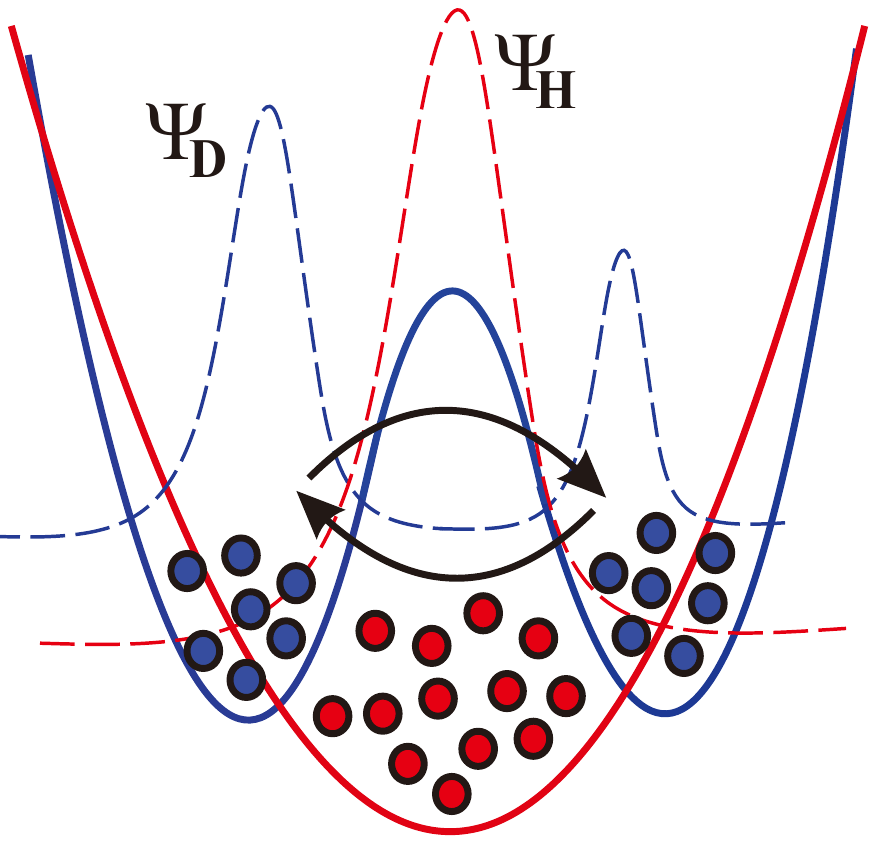}
  \caption{Schematic illustration of our setup. The solid curves illustrate the doubel-well (blue) and the harmonicl (red) potentials, respectively. The dashed curves illustrate the wave functions of the two BECs.}
  \label{fig:doublewell}
\end{figure}

We consider a BEC mixture consisting of a BEC in a deep double-well potential, with the other BEC trapped in a harmonic potential symmetrically positioned at the center of the double well  (Fig~\ref{fig:doublewell}). For simplicity, we consider a one-dimensional setup. The coupled time-dependent BECs wave function is described by the following mean-field GP equations.
\begin{align}
   i\hbar\frac{\partial\psi_D}{\partial t}=&-\frac{1}{2m}\frac{\partial^2\psi_D}{\partial x^2}+\big[V_D+g|\psi_D|^2+2g|\psi_H|^2\big]\psi_D \label{equation:1}\\
   i\hbar\frac{\partial\psi_H}{\partial t}=&-\frac{1}{2m}\frac{\partial^2\psi_H}{\partial x^2}+\big[V_H+g|\psi_H|^2+2g|\psi_D|^2\big]\psi_H
\end{align}
Here we assume the different BECs have the same atom mass $m$. respectively, $V_D=m\omega^2 x^2/2+de^{-x^2/\sigma}$ is the double-well trap and  $V_H=m\omega^2 x^2/2$ is the harmonic trap,  $d$ characterizes the barrier height of the double-well potential, $\sigma$ characterizes the width of the Guassian potential barrier, and $\omega$ is the trapping frequency of the harmonic trap. We assume that the two BECs share the same scattering length with $a_1=a_2=a_{12}, g=4\pi\hbar^2 a_1/m$, $a_{1,2}$ $(a_{12})$ is the intra (inter)-specific scattering length. (for simplicity, in the following we set $\hbar=m=\omega=1$)

The coupled GP equations above can be solved numerically. Alternatively, to simplify calculations and to gain further insight of the system dynamics, we adopt the two-mode approximation for the BEC in the double-well potential. This amounts to write
\begin{equation}
     \psi_D=b_1(t)\psi_1(x)+b_2(t)\psi_2(x) \label{equation:3}
\end{equation}
where the spatial mode functions $\psi_{1,2}(x)$ are assumed to be real and satisfy the orthonormal condition

\begin{align*}
   \int dx\psi_i(x)\psi_j(x)=\delta_{ij}
\end{align*}
which localized in each wells, the coefficients $b_{1,2}(t)=\sqrt{N_{1,2}(t)}e^{i\theta_{1,2}(t)}$, $|b_{1,2}(t)|^2=N_{1,2}(t),N_1(t)+N_2(t)=N_D$, where $N_{1,2}(t)$ is the atom number in each well, $N_D$ is the total atoms in the double-well, and $\theta_{1,2}(t)$ is the time-dependent phase in each well.
inserting decomposition~(\ref{equation:3}) into Eq.~(\ref{equation:1}), we get
\begin{align}
   i\dot b_1=(E_1+U_1|b_1|+U_{DH}^1)b_1-kb_2\\
   i\dot b_2=(E_2+U_2|b_2|+U_{DH}^2)b_2-kb_1
\end{align}
where we have defined
\begin{align}
   &E_i=\int dx\psi_i\frac{\partial^2\psi_i}{\partial x^2}+\int dx|\psi_i|^2V_D\quad i=1,2\\
   &U_i=g\int |\psi_i(x)|^4dx\\
   &U_{DH}^i=2g\int |\psi_H(x)|^2|\psi_i(x)|^2dx \\
   &k=\int dx\Big(\frac{1}{2}\psi_1\frac{\partial^2\psi_2}{\partial x^2}-\psi_1V_D(x)\psi_2
   -2g\psi_1|\psi_H|^2\psi_2\Big)
\end{align}
Here we set $d=5$, $\sigma=2.25$.
Defining the population imbalance $ Z_D(t)=\frac{(N_1-N_2)}{N_1+N_2}$ and the relative phase $\phi_D(t) =\theta_2-\theta_1$, Note that we have used the symmetric double well with $V_D(x)=V_D(-x)$, so $\psi_1(x)=\psi_2(-x)$, this make  $E_1-E_2=0, U_1=U_2=U_D$, we have
\begin{align}
   \dot Z_D=&-2k\sqrt{1-Z_D^2}\sin\phi_D \label{equation:z}\\
   \dot{\phi_D}=&U_DN_DZ_D+2k\frac{Z_D}{\sqrt{1-Z_D^2}}\cos\phi_D+(U_{DH}^1-U_{DH}^2)\label{equation:phi}\\
   i\frac{\partial \psi_H}{\partial t}=&\Big[-\frac{1}{2}\frac{\partial^2}{\partial x^2}+V_H(x)+g|\psi_H|^2+g(1+Z_D)N_D|\psi_1|^2\nonumber\\
   &+g(1-Z_D)N_D|\psi_2|^2\Big]\psi_H.\label{equation:gpe}
\end{align}
and the wave function of harmonically trapped  BEC satisfy the conservation $N_H=\int dx|\psi_H(x)|^2$. where $N_H$ is the total atoms in harmonic trap, In the limit where the atom-number density of the harmonically trapped BEC is vanishingly small, $\psi_H$ drops out of the coupled equations, and the equations above reduce to the case of the simple tunneling dynamics of a two-mode BEC. On the other hand, when the number density of the harmonically trapped BEC is overwhelmingly large, such that its self-energy is much larger than the inter-species interaction energy, the harmonically trapped BEC would hardly be affected by the tunneling dynamics, and the inter-species interaction can be approximated by a time-independent potential. The dynamics is again reduced to the tunneling dynamics of a simple two-mode BEC, albeit with the barrier potential inbetween dressed by the harmonically trapped BEC. For intermediate densities, however, we need to solve the coupled equations numerically to determine the system dynamics.

\begin{figure*}[tbp!]
  \centering
  \includegraphics[width=16.5cm,height=7cm]{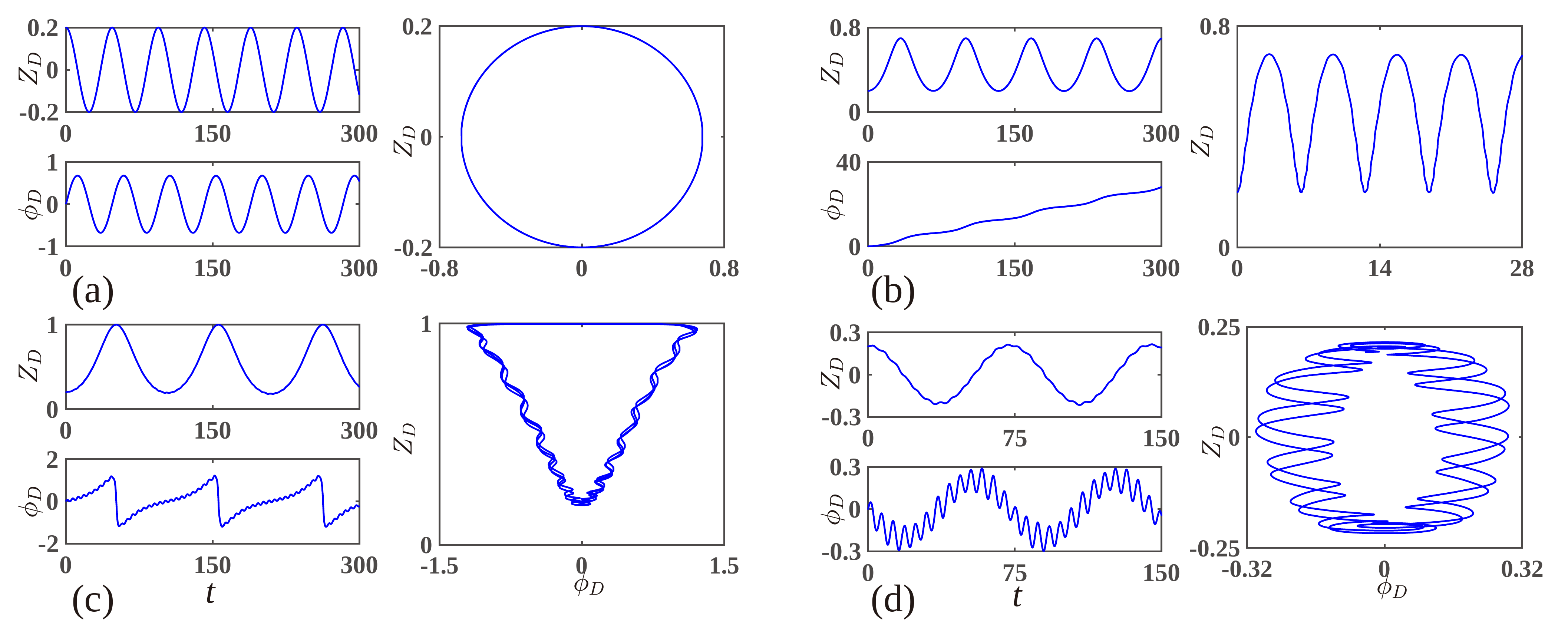}
  \caption{For zero mode, population imbalance change with time $t$, phase change with $t$ and population imbalance change with the phase. for all figures the initial condition is set $Z_D(0)=0.2,  \phi_D(0)=0$, $g=3.3\times10^{-3}$, for (a) $N_H=0$, (b) $N_H=2N_D$, (c) $N_H=3N_D$, and  (d) $N_H=4N_D$.}
  \label{fig:zeromode}
\end{figure*}

\section{Tunneling Dynamics in the BEC mixture}
For the simple tunneling dynamics of a BEC in a double-well potential, depending on the initial population imbalance $Z_D$ and relative phase $\phi_D$, the system can exhibit one of the following dynamics: the zero-phase oscillation, in which both the time average of the relative phase and the population imbalance are zero, with $\langle \phi_D(t)\rangle=0$ and $\langle Z_D(t)\rangle=0$; the $\pi$-phase oscillation with $\langle \phi_D(t)\rangle=\pi$ and $\langle Z_D(t)\rangle=0$; the MQST phase with $\langle Z_D(t)\rangle\neq 0$. As we will show from our numerical results, the back action of the harmonically trapped BEC enriches the non-linear tunneling dynamics. The resulting dynamics largely depend on the initial condition. In the following, we will discuss how the different dynamic phases are modified by increasing $N_H$, the atom number of harmonically trapped BEC.
\subsection{Zero-phase mode}
We first study the effect of $N_H$ on the zero-mode phase by setting the initial conditions as: $Z_D(0)=0.2, \phi_D(0)=0, N_D = 260, g=3.3\times 10^{-3}$. The results are shown in Fig.~\ref{fig:zeromode}(a)-(d), where
the dynamics of the population imbalance $Z_D(t)$ and the relative phase $\phi_D(t)$ are plotted along with the phase portrait of the dynamics on the $Z_D$-$\phi_D$ plane.

Figure~\ref{fig:zeromode}(a) describes the case for $N_H =0$ and shows sinusoidal oscillations, which are often referred to as plasma oscillations in analogy to the superconducting Josephson junctions. After sweeping $N_H$ equal to $2N_D$, as in Fig.~\ref{fig:zeromode}(b), the system enters the regime of self-trapping, Increasing $N_H$ would drive the system into the MQST state beyond the critical value $N_{Hc} =1.1 N_D$. In the MQST state, the tunneling is strongly suppressed, which leads to self-trapped, nearly stationary modes localized inside a single well ($\langle Z_D(t)\rangle\neq 0$). When $N_H$ is only slightly beyond $N_{Hc}$, the relative phase begins to vary monotonically with time as shown in Fig~\ref{fig:zeromode}(b). As the evoluation of the relative phase is unbounded in this MQST state, we call it the running-phase MQST, which we label as MQST$_1$. On increasing $N_H$ further to $3N_D$ , the system remains self-trapped, however, the evolution of the relative phase becomes bounded as well, as shown in Fig~\ref{fig:zeromode}(c). We call the MQST state here the bounded MQST, which we label as MQST$_2$. Finally, as $N_H$ becomes very large, the system dynamics should recover the Josephson-like oscillations. In Fig~\ref{fig:zeromode}(d), we see that the dynamics with $N_H=4N_D$ is already qualitatively similar to those in Fig~\ref{fig:zeromode}(a), although sizeable anharmonics can still be identified due to the finiteness of $N_H$.

\subsection{$\pi$-phase mode}
\begin{figure*}[tbp]
  \centering
  \includegraphics[width=16.5cm,height=7cm]{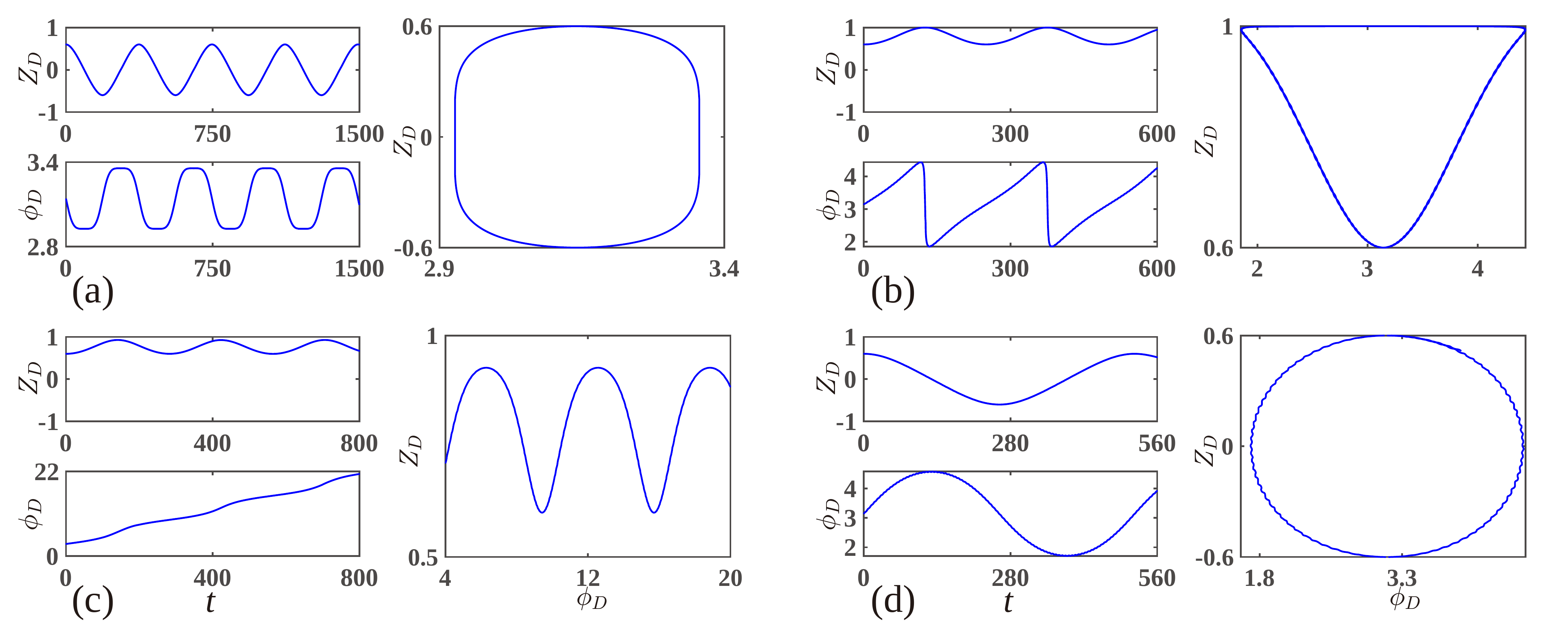}
  \caption{For $\pi$- mode, population imbalance change with time $t$, phase change with $t$ and Population imbalance change with the phase. for all figures the initial condition is set $Z_D(0)=0.6,  \phi_D(0)=\pi$, $g=4\times10^{-4}$, for (a) $N_H=0$, (b) $N_H=8N_D$, (c) $N_H=10N_D$, and  (d) $N_H=14N_D$.}
  \label{fig:pimode}
\end{figure*}

We then study the effect of $N_H$ on the $\pi$-mode phase by setting the initial conditions as: $Z_D(0)=0.6, \phi_D(0)=\pi, N_D = 210, g=4\times 10^{-4}$.  The results are shown in Fig.~\ref{fig:pimode}(a)-(d), where
the dynamics of the population imbalance $Z_D(t)$ and the relative phase $\phi_D(t)$ are plotted along with the phase portrait of the dynamics on the $Z_D$-$\phi_D$ plane. Similar to the previous case, by increasing $N_H$ gradually, the system dynamics change through different regimes: first the Josephson oscillation, then the bounded MQST (MQST$_2$), then the running-phase MQST (MQST$_1$), and finally the Josephson-like oscillation.  Notably, for intermediate $N_H$ here, the system dynamics enter the MQST$_2$ first, with bounded $Z_D$ and $\phi_D$, before entering the MQST$_1$ regime. In the large $N_H$ limit, the system again recovers the $\pi$-mode Josephson oscillation.

\subsection{MQST}
\begin{figure*}
\centering
\includegraphics[width=16.5cm,height=7cm]{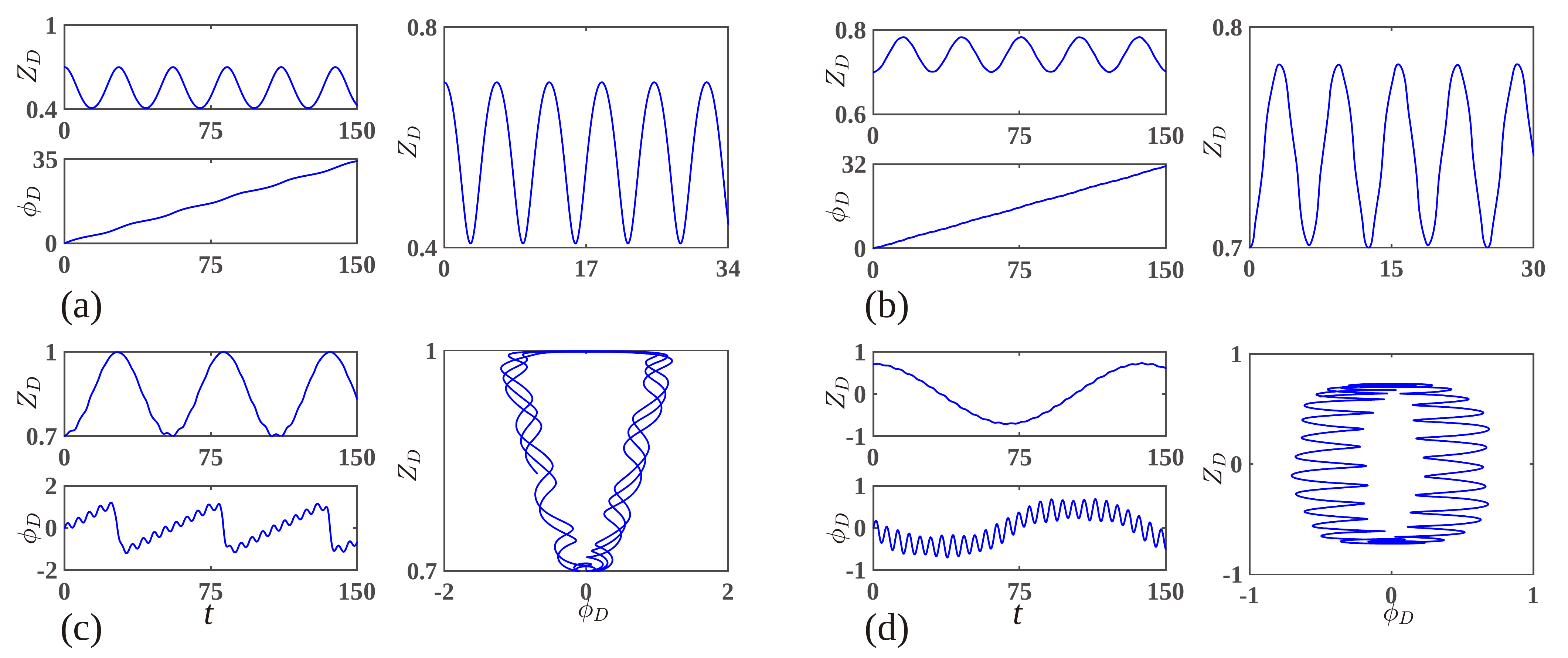}
\caption{For running mode, population imbalance change with time $t$, phase change with $t$ and Population imbalance change with the phase. for all figures the initial condition is set $Z_D(0)=0.7$, $\phi_D(0)=0$, $g=3.3\times10^{-3}$, for (a) $N_H=0$, (b) $N_H=1.5N_D$, (c) $N_H=2.8N_D$, (d) $N_H=3.5N_D$}
  \label{fig:runningmode}
  \end{figure*}

Finally, we study the effect of $N_H$ on the MQST phase by setting the initial conditions: $Z_D(0)=0.7$, $\phi_D(0)=0, N_D = 260$, and $g = 3.3 \times 10^{-3}$. As illustrated in Fig.~\ref{fig:runningmode}, when $N_H$ increases, the dynamics changes from the running-phase MQST with  $\langle Z_D(t)\rangle< Z_D (0)$, to a running-phase MQST with $ \langle Z_D(t)\rangle> Z_D (0)$. On further increasing $N_H$, the system enters bounded MQST, and finally to a zero-phase-like oscillating mode, which suggests that the dynamics of the double-well BEC recovers the Josephson-like oscillation in the large $N_H$ limit. In Fig.~\ref{fig:runningmode}, we only show the system dynamics starting from the running phase MQST. The dynamics starting from the bounded MQST are qualitatively the same as shown in Fig.~\ref{fig:pimode}(e)(f).

\begin{figure*}[tbp!]
  \centering
  \includegraphics[width=15cm,height=4cm]{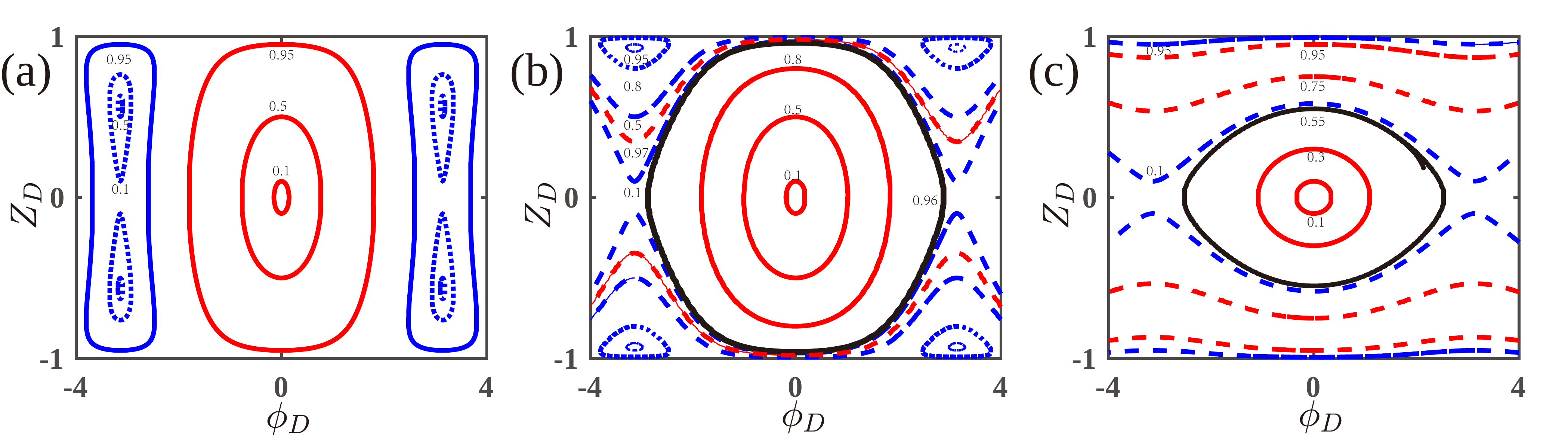}
  \caption{ Phase-portrait of the bosonic Josephson junction in the two mode approximation for different number of harmonically trapped atoms $N_H$, (a) $N_H=0$, (b) $N_H=6N_D$, (c) $N_H=8.8N_D$. Trajectories with $\phi_D(0) = 0$ are depicted in red while those with $\phi_D(0) =\pi$ are depicted in blue. The dashed lines represent the running phase MQST while the dotted lines correspond to bounded MQST. The indicated numbers represent the different initial population imbalances $Z_D(0)$. The black lines in (b) and (c) depict the respective separatrix, which is the phase-plane trajectory for $Z_D(0) = Z_{Dc}$.}
  \label{fig:PPP}
\end{figure*}

\section{Phase-plane portrait and phase diagram}

So far, we have been studying the effects of increasing $N_H$ on the system dynamics for given initial parameters. To establish an overall picture of the system dynamics, it is helpful to systematically examine the phase portrait of the system, i.e., the trajectory of the system parameters on the $Z_D$-$\phi_D$ plane. In Fig.~\ref{fig:PPP}, we show such portraits with different initial conditions and for fixed values of $N_H$ in each subplot. In these figures, the various regimes of the nonrigid pendulum dynamics discussed above can be summarized very intuitively in terms of different constant-energy contours.
\begin{figure}[tbp!]
  \centering
  \includegraphics[width=7cm,height=10cm]{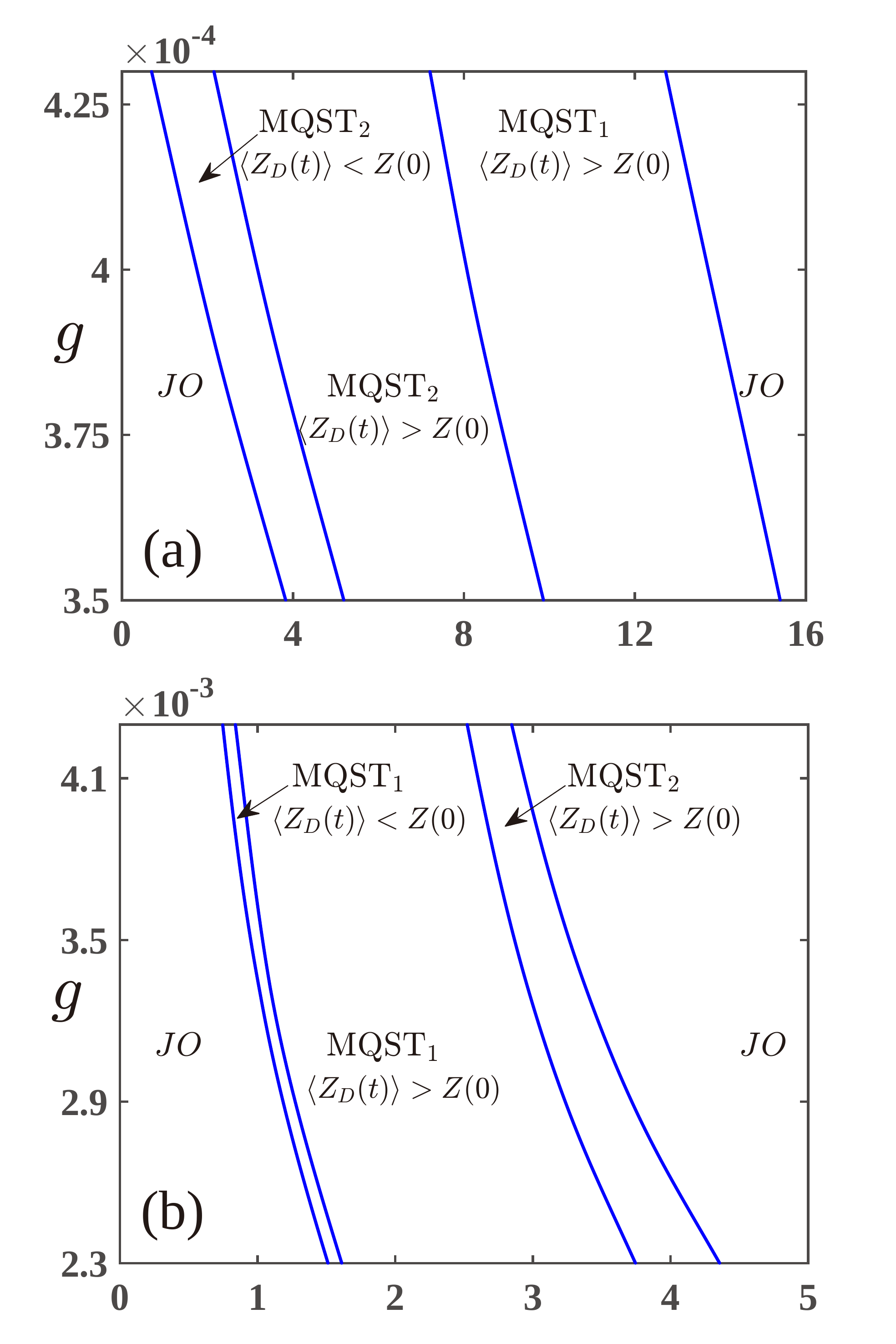}
  \centerline{$N_H/N_D$}
  \caption{Phase diagram for $\pi$ mode (a) and for zero mode (b) representing different regimes and corresponding widths of each region.}
  \label{fig:ppd}
\end{figure}
\begin{figure}[tbp!]
  \centering
  \includegraphics[width=7cm,height=10cm]{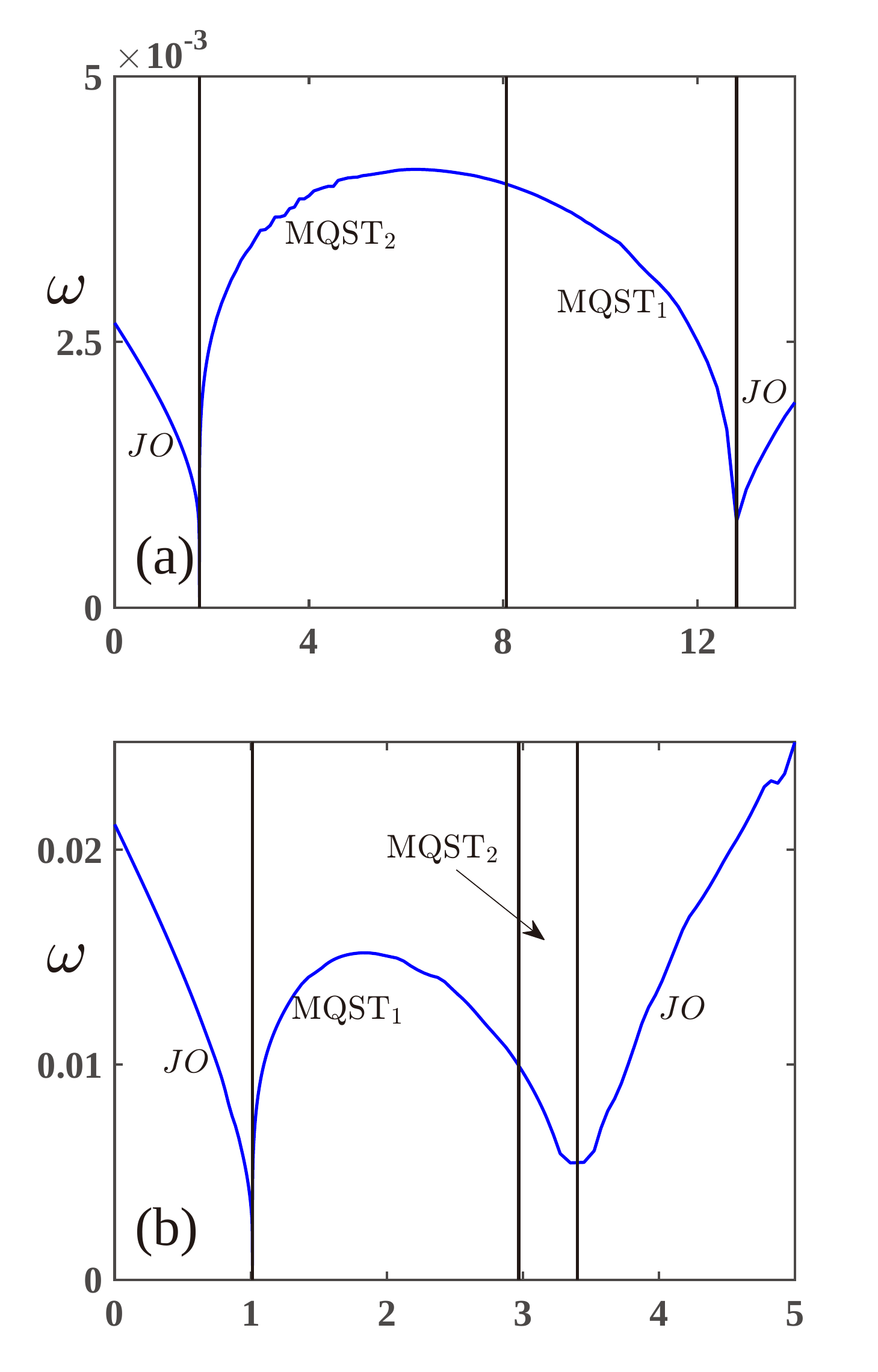}
  \centerline{$N_H/N_D$}
  \caption{
The double well BEC frequency $\omega$ change with the atom number in the harmonic trap. for figure (a) $Z_D=0.6$, $\phi_D=\pi$, $N_D=210, g=4\times 10^{-4}$. (b) $Z_D=0.2$, $\phi_D=0$, $N_D=260, g=3.3\times 10^{-3}$.}
  \label{fig:OFC}
\end{figure}

\begin{figure}[tbp!]
  \centering
  \includegraphics[width=8.5cm,height=9.56cm]{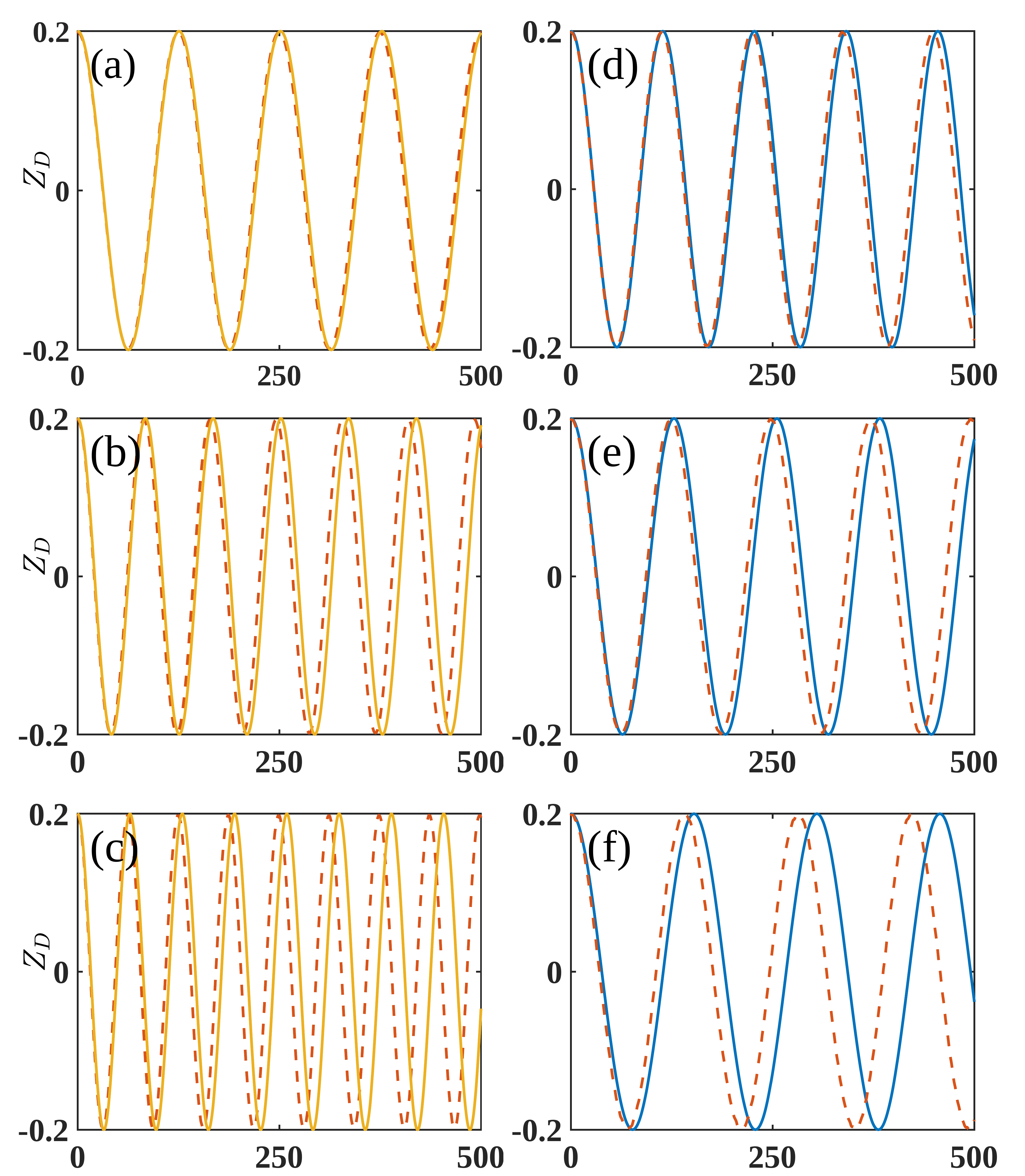}
  \centerline{{\bf $t$}}
  \caption{The difference between two-mode model and GP equations, for figure (a),(b),(c) fix the $N_H=0, g=4\times 10^{-4}$, and change $N_D=100,500,1000$, for figure (d),(e),(f) we fix the $N_D=200, g=4\times 10^{-4}$, and change the $N_H$ from $\frac{1}{2}N_D$, $2N_D$ to $4N_D$, solid line is the results from two-mode model and the dash-dot line is the GP equations results.}
  \label{fig:TM}
\end{figure}
Fig~\ref{fig:PPP}(a) shows the cases with $N_H = 0$, where we have colored zero-phase, $\pi$-phase, and dashed  MQST modes differently. Here, no running-phase MQST is present, and the system can enter a bounded MQST for  $Z_{D}(0)<Z_{Dc} \sim 0.77$.
Fig ~\ref{fig:PPP}(b) shows the cases with $N_H = 6N_D$. In this case, a transition from the zero-phase mode to the running-phase MQST can be identified at the critical value $Z_{Dc}\sim0.96$. This separatrix is indicated with a thick black line. In Fig ~\ref{fig:PPP}(b), for an initial phase $\phi_D (0) = \pi$, all trajectories now become self-trapped (as opposed to Fig~\ref{fig:PPP}(a)). The transition from the running-phase MQST to the $\pi$-phase MQST happens at $Z_{Dc}\sim0.67$. Finally, Fig ~\ref{fig:PPP}(c) shows the various regimes for $N_H = 8.8N_D$. While the regime for the zero-phase mode further decreases, the rest of the phase portrait is dominated by the running-phase MQST. The transition between the zero-phase mode and the running-phase MQST occurs around $Z_{Dc}\sim0.55$.

From the phase portrait, it is clear that an important effect of the harmonically trapped BEC on the dynamics of the double-well BEC is the enhancement of the MQST, and particularly the running-phase MQST, at the cost of Josephson-like oscillations. This can be further confirmed by looking at the phase diagram of the system. Figure~\ref{fig:ppd} shows the phase diagram for system initially in the zero- or $\pi$-phase- oscillation regime at $N_H=0$. Apparently, inbetween the small and the large $N_H$ regimes, the system dynamics are governed by the running-phase MQST (MQST$_1$) or the bounded MQST (MQST$_2$) modes.
\section{Oscillation Frequency and detection}
We now study the oscillation frequency of the double-well BEC in response to the variation of $N_H$. In Fig.~\ref{fig:OFC}, we show how the frequency changes with increasing $N_H$. It is apparent that between regimes of Josephson-like oscillations and those of the MQST modes, the frequency undergoes sudden changes, suggesting a dynamic phase transition. For systems starting from a typical $\pi$-mode oscillation at $N_H=0$ (Fig.~\ref{fig:OFC}(a), when we increase $N_H$, the oscillation frequency decreases and becomes zero near $N_H\sim 1.73N_D$, when the system enters the MQST$_1$ regime. A similar kink exists for sufficiently large $N_H$ near $N_H\sim 12.8N_D$, when the system dynamics changes from MQST$_2$ regime into the Josephson-like oscillations.
In comparison, for systems starting from a typical zero-mode phase at $N_H=0$ (Fig.~\ref{fig:OFC}b), while the overall picture of the frequency change is similar, the MQST regions are now much narrower. Importantly, the critical changes of the oscillation frequency of the double-well BEC serve as signals for the phase transitions between the Josephson-oscillation dynamics and the MQSTs.

\section{Difference between two-mode models and GP equations}
We now check the validity of the two-mode approximation by comparing results with those from the coupled GP equations. As shown in Fig.~\ref{fig:TM}, the two-mode approximation is good when neither $N_H$ or $N_D$ is large. In the large-$N_D$ limit, two-mode approximation already breaks down due to inter-well interactions terms that are neglected in the current formalism. In the large-$N_H$ limit, on the other hand, the oscillation of the BEC in the harmonic trap would induce asymmetry in the local wave functions of the two wells, which effectively couples the BEC in the double-well to high-lying states of the double-well potential, and makes the two-mode approximation invalid. Therefore, our previous numerical results should be accurate when $N_H$ and $N_D$ are not too large. For larger $N_H$ and $N_D$, we expect that our results serve as a qualitatively correct description on the variation of the system dynamics.

\section{Conclusion}
We study the interaction-modulated tunneling dynamics of in a two-mode BEC coupled to be background BEC. Adopting a two-mode approximation for the BEC in the double-well potential and coupling it with the GP equation of the harmonically trapped BEC, we numerically investigate the coupled dynamics of the BEC mixture. We focus on back action of the BEC in the harmonic trap, and demonstrate strong non-linearity in the oscillation dynamics of the two-mode BEC. In particular, we show that the self-trapping mode is strongly enhanced by the harmonically trapped BEC. The dynamic phase transition between the Josephson-like oscillation and the self-trapping mode can be probed by monitoring the oscillation frequency of the double-well BEC when the interaction energy between the two BECs are tuned.

Acknowledgements:We thank Wei Zhang, Wu Fan for helpful discussions. This work is supported by the National Key R\&D Program (Grant No. 2016YFA0301700), the National Natural Science Foundation of China (Grant Nos. 60921091, 11374283, 11522545). M.Maraj acknowledges the support from CAS-TWAS president fellowship, WY acknowledges support from the ``Strategic Priority Research Program(B)'' of the Chinese Academy of Sciences, Grant No. XDB01030200.

\end{document}